\documentclass[pre, twocolumn, amsmath, amssymb, superscriptaddress]{revtex4}

\usepackage{graphicx}% Include figure files
\usepackage{dcolumn}% Align table columns on decimal point
\usepackage{bm}% bold math
\usepackage{braket}
\usepackage{placeins}
\usepackage{balance}
\usepackage{xcolor}

\usepackage{amsmath}

\newcommand{\beq}{\begin{equation}}
\newcommand{\eeq}{\end{equation}}

\begin{document}

\title{Far-Infrared Signatures for a Two-Steps Pressure-Driven Metallization in Transition Metal Dichalcogenides }

\author{Elena Stellino}
\affiliation
{University of Perugia, Department of Physics and Geology, via Alessandro Pascoli, Perugia, Italy}

\author{Beatrice D'Alò}
\affiliation
{Sapienza University of Rome, Department of Physics, P.le A. Moro 5, Rome, Italy}

\author{Francesco Capitani\footnote{Corresponding author: francesco.capitani@synchrotron-soleil.fr}
}
\affiliation
{Synchrotron SOLEIL, L’Orme des Merisiers, Saint-Aubin, Gif-sur-Yvette, France}

\author{Marine Verseils}
\affiliation
{Synchrotron SOLEIL, L’Orme des Merisiers, Saint-Aubin, Gif-sur-Yvette, France}

\author{Jean-Blaise Brubach}
\affiliation
{Synchrotron SOLEIL, L’Orme des Merisiers, Saint-Aubin, Gif-sur-Yvette, France}

\author{Pascale Roy}
\affiliation
{Synchrotron SOLEIL, L’Orme des Merisiers, Saint-Aubin, Gif-sur-Yvette, France}

\author{Alessandro Nucara}
\affiliation
{Sapienza University of Rome, Department of Physics, P.le A. Moro 5, Rome, Italy}

\author{Caterina Petrillo}
\affiliation
{University of Perugia, Department of Physics and Geology, via Alessandro Pascoli, Perugia, Italy}

\author{Paolo Postorino}
\affiliation
{Sapienza University of Rome, Department of Physics, P.le A. Moro 5, Rome, Italy}

\begin{abstract}
We present a high-pressure investigation of the semiconductor-to-metal transition in MoS$_2$ and WS$_2$ carried out by synchrotron-based far-infrared spectroscopy, to reconcile the controversial estimates of the metallization pressure found in the literature and gain new insight into the mechanisms ruling this electronic transition. Two spectral descriptors are found indicative of the onset of metallicity and of the origin of the free carriers in the metallic state:  the absorbance spectral weight, whose abrupt increase defines the metallization pressure threshold, and the asymmetric lineshape of the E$_{1u}$ peak, whose pressure evolution, interpreted within  the Fano model, suggests the electrons  in the metallic state originate from n-type doping levels. Combining our results with those reported in the literature, we hypothesize a two-step mechanism is at work in the metallization process, in which the pressure-induced hybridization between doping and conduction band states drives an early metallic behaviour, while the band-gap closes at higher pressures.
\end{abstract}

\maketitle

\section{Introduction}
Transition Metal Dichalcogenides (TMDs) are layered crystals that exhibit peculiar physical and chemical properties of great technological interest \cite{Manzeli2017, Choi2017, Mak2016}. These are mainly related to their crystalline structure made up of the stacking of weakly interacting planes. Planes are held together by strong covalent bonds while the interaction among them is mainly due to weak Van der Waals (VdW) forces. These materials are, thus, strongly anisotropic with respect to their mechanical, electrical and optical properties \cite{Johari2011, Kumar2012}.
\\
The layered structure of TMDs makes the inter-layer interaction a key parameter governing their physical properties, from the lattice dynamics to the electronic structure \cite{Zhang2015, Sun2016, Yun2012}. In this perspective, the application of pressure provides an effective way to directly modulate the weak VdW forces linking adjacent layers, allowing getting a deeper insight into the fundamental physics of these materials and paving the way for a wide range of technological applications, in which the design of electronic devices with tailored features is of utmost interest \cite{Wang2015, Wang2012}. 
\\
Several pressure-dependent studies have been carried out aiming at studying the TMD response to a strong enhancement of the inter-layer interaction \cite{Dave2004, Bhattacharyya2012, PimentaMartins2022, Guo2013}. Most of the semiconducting crystals in the bulk form were found to undergo a pressure-induced transition toward a metallic state \cite{Fan2016, Zhao2019, Stellino2021}. However, in many cases, a certain degree of ambiguity  persists in the exact determination of the metallization pressure as well as in the explanation of the microscopic origin of the transition. 
\\
In bulk MoTe$_2$, resistivity measurements found evidence of the onset of a metallic behavior at $P_M^{res} \sim$10-12 GPa \cite{Zhao2019, Yang2019}, while the effective closure of the band gap was observed at $\sim$ 24 GPa through absorption measurements in the near infrared (NIR) \cite{Stellino2022}. In a similar vein, resistivity  \cite{Nayak2014} and impedance  \cite{Zhuang2017} measurements carried out on bulk MoS$_2$ suggested the crystal becomes metallic at $P_M^{res} \sim$20 GPa, while absorption measurements in the  NIR/visible range showed that the band gap closes at pressures higher than 26 GPa \cite{BrotonsGisbert2018}. As for the other semiconductors belonging to the class of TMDs, like WS$_2$, in the majority of cases, the definition of the metallization threshold only relies on transport measurements \cite{Nayak2015}, while there lacks a direct estimate of the pressure-dependent trend of the band gap.
\\\\
In the present work, we investigate the mechanisms at the origin  of the metallic transition in semiconducting TMDs analysing, by synchrotron-based far-infrared (FIR) spectroscopy, the pressure-induced changes in the optical response of MoS$_2$ and WS$_2$. We found that the FIR spectrum can provide us with two effective descriptors that indicate both the onset of the metallic behaviour and the origin of the free charge carriers formed in the process. The first one is the absorbance spectral weight \cite{Stellino2021, Stellino2022, Postorino2003, Calvani1998}, whose abrupt increase under pressure defines the metallization threshold, $P_M^{FIR}$, in good agreement with the results from resistivity measurements found in the literature for both compounds (in MoS$_2$, $P_M^{res}\sim$20 GPa \cite{Nayak2014} and $P_M^{FIR}\sim$ 20 GPa, in WS$_2$, $P_M^{res}\sim$22 GPa \cite{Nayak2015} and $P_M^{FIR}\sim$ 21 GPa). The second one is the lineshape of the E$_{1u}$ phonon, whose pressure evolution, interpreted in the framework of the Fano theory \cite{Fano1961}, suggests a reliable explanation for the origin of the free charge carriers in the metallic state. The asymmetric profile shown by the E$_{1u}$ phonon at ambient conditions is symptomatic of the coupling between the vibrational mode and a continuum of electronic transitions with comparable energy scales. In both MoS$_2$ and WS$_2$, these transitions cannot originate from a standard excitation from valence to conduction band, which is far too energetic ($\sim$ eV) for the phonon ($\sim$10$^{-2}$ eV) to couple with, but they reasonably occur from doping levels located in the proximity of the conduction band minimum. Our measurements on MoS$_2$ and WS$_2$ show that, in correspondence with the increase in the absorbance spectral weight, the E$_{1u}$ peak abruptly symmetrizes, suggesting the energy separation between the doping levels and the conduction band minimum goes to zero. 
\\
Based on this phenomenology, our hypothesis is that the metallic behaviour we observed at $P_M^{FIR}\sim$20 GPa and $P_M^{FIR}\sim$21 GPa for MoS$_2$ and WS$_2$ respectively is driven by the injection of doping electrons in the conduction band, while the effective closure of the indirect band gap occurs at higher pressure, as indicated by NIR/visible optical measurements. 
\\
The pressure evolution of the MoS$_2$ and WS$_2$ absorbance spectra closely resembles that of MoTe$_2$, previously studied and explained within the aforementioned scheme \cite{Stellino2022}. The observation of a similar phenomenology in three distinct compounds with three different metallization pressures might indicate that the observed behaviour is a common feature of semiconducting TMDs, suggesting the exploitation of our interpretation model in the general description of the metallic transition of these compounds.

\section{Experimental}
MoS$_2$ and WS$_2$ single crystals were provided by HQ-Graphene. Measurements were performed on fresh-cut samples, directly cleaved from the macroscopic crystal to obtain flakes a few microns thick with a clean surface. It is worth noticing that, since the typical thickness of single-layer TMDs is $\sim$ 0.7 nm, in the present case, we safely work in the bulk limit.
\\\\
Room temperature infrared transmission measurements on MoS$_2$ were performed at the beamline SMIS of the SOLEIL synchrotron over the 0-26 GPa range. In the DAC, diamonds with 250 $\mu$m culet were separated by a pre-indented, stainless steel, 50 $\mu$m thick gasket, in which a hole of 125 $\mu$m in diameter was drilled. Measurements were performed using a Thermo Fisher iS50 interferometer equipped with a solid-substrate beamsplitter. Synchrotron edge radiation was employed as a light source.
Custom-made 15x Cassegrain objectives with a large working distance allowed focusing and then collecting the transmitted radiation, finally detected by a liquid-helium-cooled bolometer detector.
\\
Infrared transmission measurements on WS$_2$ were performed at the beamline AILES of the SOLEIL synchrotron over the 0-25 GPa range. In the Diamond Anvil Cell (DAC), diamonds with 400 $\mu$m culets were separated by a pre-indented stainless steel 50 $\mu$m thick gasket, in which a hole of 150 $\mu$m in diameter was drilled. Measurements were carried out using a Bruker IFS 125 HR interferometer, coupled with the synchrotron source, equipped with  15x Cassegrain objectives, a multi-layer Mylar 6 $\mu$m beamsplitter and a liquid Helium cooled bolometer. 
\\
Both samples were positioned in the hole together with polyethylene (for WS$_2$) or CsI (for MoS$_2$) as a pressure transmitting media and a ruby chip, to measure the pressure through the ruby fluorescence technique.
\\
At each pressure, the absorbance spectrum $A(\tilde \nu)$ was obtained as \\$A(\tilde \nu)=-ln[I(\tilde \nu)/I_0(\tilde \nu)]$, where $I(\tilde \nu)$ is the intensity transmitted with the sample loaded in the cell and $I_0(\tilde \nu)$ is the background intensity with the DAC filled by the hydrostatic medium only.

\section{Evolution of the absorption spectral weight under pressure}
\begin{figure*}[t]
    \centering
    \includegraphics[width=0.6\textwidth]{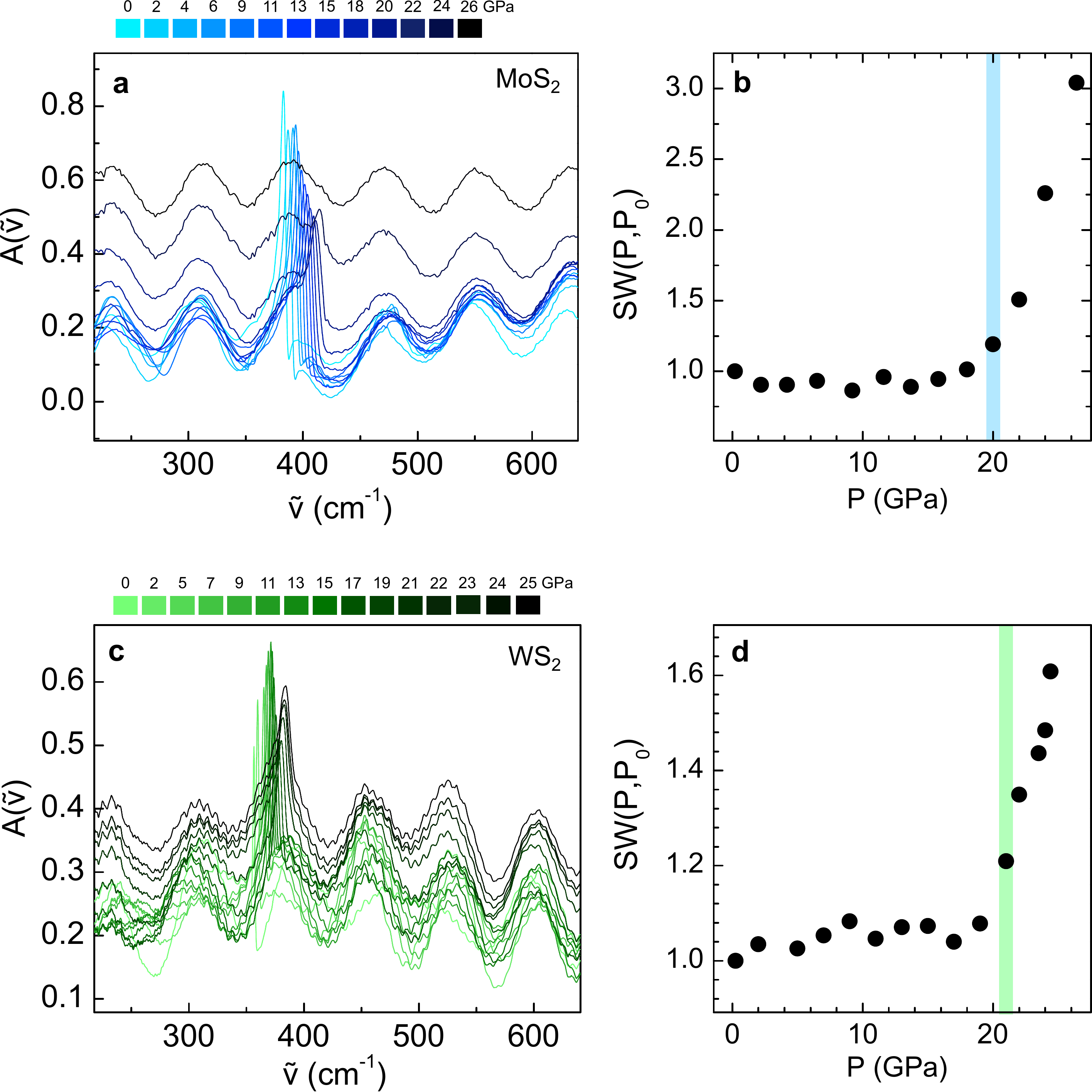}
    \caption{(a) MoS$_2$ absorbance spectra in the 0-26 GPa pressure range measured on the SMIS beamline. (b) Normalized spectral weight $SW(P, P_0)$, calculated in the 200-600 cm$^{-1}$ range, as a function of pressure. The light blue stripe indicates the pressure at which $SW(P, P_0)$ undergoes a 30$\%$ of increase with respect to the average value obtained at lower pressures. (c) WS$_2$ absorbance spectra in the 0-25 GPa pressure range measured on the AILES beamline. (b) Normalized spectral weight $SW(P, P_0)$, calculated in the 200-600 cm$^{-1}$ range, as a function of pressure. The light green stripe indicates the pressure at which $SW(P, P_0)$ undergoes a 20$\%$ of increase with respect to the average value obtained at lower pressures.}
    \label{figura_SW}
\end{figure*}

\balance
As shown in figure \ref{figura_SW}, in both samples, all the measured $A(\tilde \nu)$ curves display quite intense oscillating fringes due to interference effects between the internal diamond faces through the hydrostatic medium in $I_0(\tilde \nu)$. In this case, a  Fourier transform smoothing procedure is not effective to remove the fringes owing to the reduced number of oscillation present in the spectrum. Besides the fringes, at ambient pressure, a sharp peak is well distinguishable at $\sim$380 cm$^{-1}$ in MoS$_2$ and at $\sim$350 cm$^{-1}$ in WS$_2$, assigned to the in-plane infrared phonon E$_{1u}$. The phonon peak evolution under pressure will be discussed in detail in the next section, while, here, we will focus on the overall trend of the $A(\tilde \nu)$ spectra. As we can notice, in both MoS$_2$ (figure \ref{figura_SW} a) and WS$_2$ (figure \ref{figura_SW} c), the average $A(\tilde{\nu})$ value, outside the phonon region, remains almost constant at low pressure and, then, abruptly rises above a pressure threshold, $P^{FIR}_{M}$, around 20 GPa. In the framework of the Drude model, if $\tilde \nu/\Gamma \ll 1$, with $\Gamma=c/\tau$ and $\tau$ Drude relaxation time, the absorbance can be written in terms of the DC conductivity $\sigma_0$ and the frequency $\tilde \nu$ as $A(\tilde \nu) \propto \sqrt{\sigma_0 \tilde \nu}$. In the FIR range here explored, we see that the Drude band contributes to the spectrum as a  wavenumber-independent background, suggesting a relatively large value for $\Gamma$ compared to the  $\tilde \nu$ values considered and, thus, allowing us to work within the low-wavenumber approximation. Based on these considerations, we can say that the increase we observed in the FIR absorbance is directly related to an increase in $\sigma_0$ \cite{Stellino2021, Postorino2003, Calvani1998} and, thus, to the onset of the metallic behaviour reported in the resistivity measurements of both MoS$_2$ and WS$_2$ \cite{Nayak2014, Nayak2015}.
\\
In order to quantitatively characterize this observation, we define at each pressure the absorption spectral weight as 
$\displaystyle{sw(P)=\int_{\tilde \nu_m}^{\tilde \nu_M}{A(\tilde \nu)d\tilde \nu}}$. The trend of the normalized quantity $ SW(P,P_0)= sw(P)/sw(P_0)$, where $P_0$  

\FloatBarrier
\noindent corresponds to the minimum pressure value measured in the two samples, is reported in figures \ref{figura_SW} b,d for MoS$_2$ and WS$_2$ respectively. The integration is performed in the range 200-600 cm$^{-1}$, but it excludes the small wavenumber interval where the phonon contribution is dominant. In the MoS$_2$ case, we can notice a $\sim$30$\%$ of increase in  $SW(P,P_0)$ at 20 GPa compared with the average value calculated at lower pressures. Therefore, we can define the metallization threshold $P_{M} = 20 \pm 1$ GPa, in very good agreement with the results from resistivity measurements that observe the onset of the transition at $\sim$ 20 GPa \cite{Nayak2014}. As for the WS$_2$ crystal, a $\sim$20$\%$ of increase is observed going from the 0-20 GPa range to 21 GPa, allowing us to place the metallization threshold at $P_{M} = 21 \pm 1$ GPa. Again, this result is coherent with the transition pressure ($\sim$22 GPa) found by resistivity measurements from the literature \cite{Nayak2015}.
\\
A comparison between the spectral weight trends of the two samples reveals that the rate of increase of $SW(P,P_0)$ above $P^{FIR}_{M}$ is significantly higher in MoS$_2$ than in WS$_2$. Indeed, if we perform a linear fit of $SW(P,P_0)$ for values higher than $P^{FIR}_{M}$ we obtain a slope of $\sim$0.3 GPa$^{-1}$ in the MoS$_2$ trend and of $\sim$0.1 GPa$^{-1}$ in the WS$_2$ trend. This behaviour is consistent with the results from transport measurements on both samples \cite{Nayak2014, Nayak2015}, where the authors report an abrupt drop in the MoS$_2$ resistivity ($\rho$) in correspondence with the metallization onset, while a more gradual decrease in $\rho$ was observed in the case of WS$_2$.

\section{Evolution of the Fano resonance under pressure}
\begin{figure*}[t]
    \centering
    \includegraphics[width=0.6\textwidth]{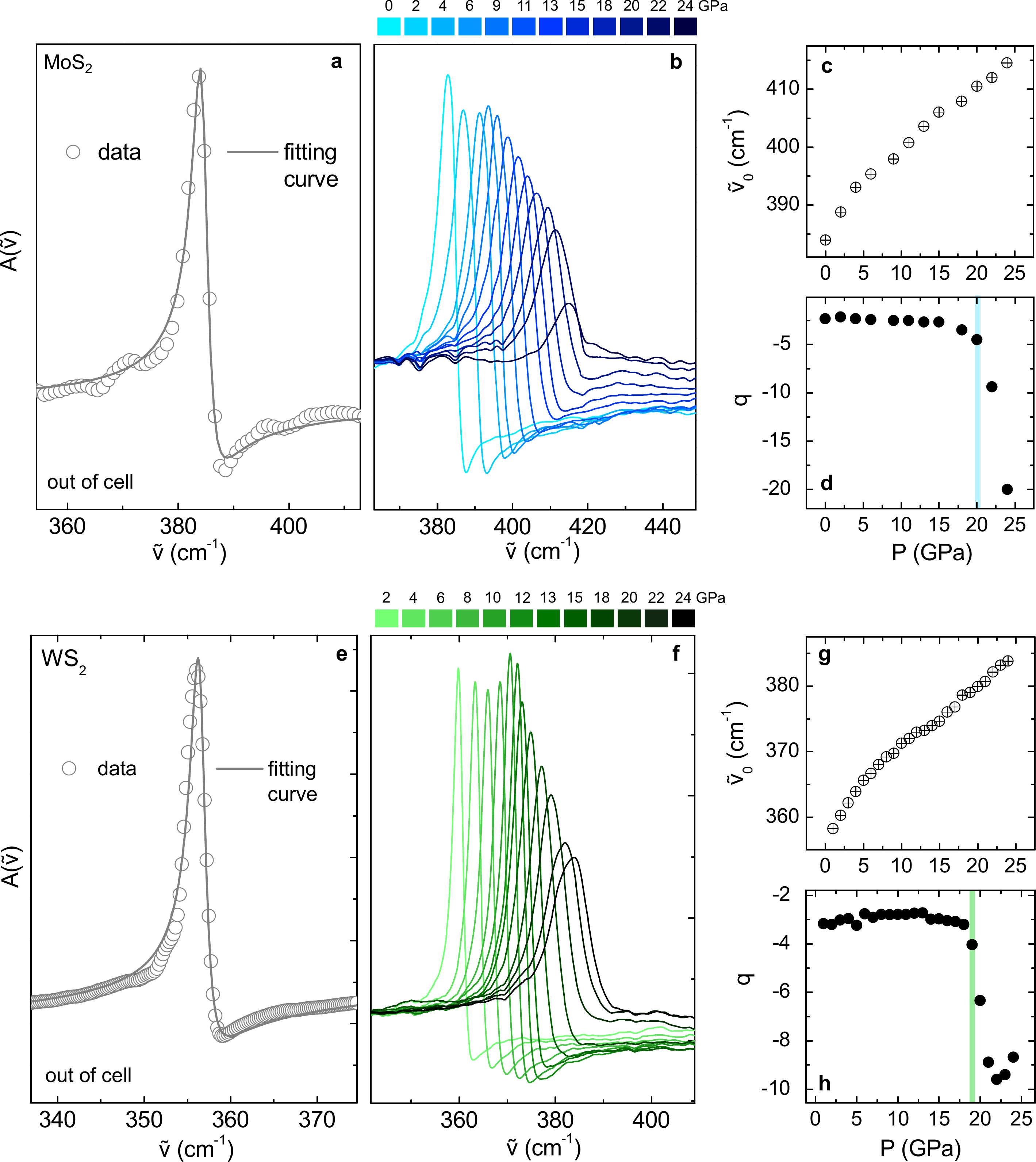}
    \caption{(a), (e) MoS$_2$, WS$_2$ absorbance spectra at selected pressures after the subtraction of the polynomial background and the sinusoidal fringes. The offsets have been shifted in order to ease the peak comparison. (b), (g) MoS$_2$, WS$_2$ absorbance spectra collected out of the DAC (dots) fitted by a Fano function (continuous line). (c), (g) E$_{1u}$ peak position $\tilde \nu_0$ as a function of pressure, in MoS$_2$ and WS$_2$ respectively. (d), (h) $q$ parameter of the Fano fit as a function of pressure, in MoS$_2$ and WS$_2$ respectively. The light blue (green) stripe indicates the pressure at which the $q$ parameter visibly starts to decrease in MoS$_2$ (WS$_2$).}
    \label{figura_fano}
\end{figure*}

\balance
 As seen in the previous section, FIR absorption spectra exhibit an intense peak at $\sim$ 380 cm$^{-1}$ in  MoS$_2$ and at $\sim$ 350 cm$^{-1}$ in WS$_2$, both attributed to E$_{1u}$ vibrational modes. Figures \ref{figura_fano}a,e show the $A(\tilde{\nu})$ spectra of MoS$_2$ and WS$_2$, respectively, collected out of the DAC, in the wavenumber region around the phonon centre. In this configuration, the absence of interference fringes allows us to carefully examine the peak lineshape, which appears to significantly deviate from a standard Lorentzian in favour of a Fano-like profile. 
 \\
 The Fano theory, first introduced to describe the anomalous profile of the absorption peaks observed in the spectrum of  He \cite{Fano1961}, describes, in general, the effects of the interaction between a discrete quantum state and a continuum of states with comparable energy scales. The model was reformulated by Rice et al. \cite{Rice1976} in terms of the so-called  \textit{charged phonon theory}, in which phonon modes with an asymmetric absorption profile are derived, under specific conditions, when the electron-phonon interaction is perturbatively inserted in the crystal Hamiltonian. In this framework, the gate-tunable asymmetric lineshape of the $E_{u}$ mode in graphene has been recently interpreted considering the coupling between the vibrational mode and the continuum formed by electron-hole transitions  \cite{Cappelluti2009, Cappelluti2012}.
\\
The close resemblance between TMDs and graphite in terms of lattice structure makes it possible, in principle, to interpret the phonon anomalies observed in the FIR spectrum of these crystals with the same model proposed for graphite, provided that we find suitable electronic excitations coupled with the phonon mode. In both MoS$_2$ and WS$_2$, the energy of \textit{standard} valence-to-conduction band  transitions is higher than $E'_g\sim$1.5-1.6 eV (that is the approximated value of the indirect gap of both samples), while the phonon energy, $E_{ph} \sim$ 40-65 meV, is nearly two orders of magnitude smaller. Therefore, a resonance between these two excitations is highly improbable. Since most of semiconducting TMDs are affected by a small percentage of n-type doping ( $10^{14}-10^{15}$ cm$^{-3}$, as reported by the manufacturer HQ-Graphene), our hypothesis is that the electrons in excess may produce some extra levels right below the conduction band \cite{Leonhardt2020, Grzeszczyk2021, McCreary2016, Pisoni2015}. The thermal excitations from these levels to the empty states in the conduction band could be of the same energy scale of the phonon (indeed, $K_B T \sim 25$ meV and the phonon energies are lower than $50$ meV), providing the electron-phonon coupling mechanism responsible for the Fano lineshape observed in the absorption peaks. 
\\
Following this interpretation, we apply a fitting procedure for the phonon spectra of MoS$_2$ and WS$_2$ at ambient conditions using the Fano function reported here:
\begin{equation} \label{eq}
F(\omega)=F\cdot\frac{q^2+2q(2(\tilde{\nu}-\tilde{\nu}_0)/ \gamma) -1}{q^2\cdot(1+(2(\tilde{\nu}-\tilde{\nu}_0)/ \gamma)^2)}
\end{equation}

where $\tilde{\nu}_0$ is the peak centre, $\gamma$ is the peak width and $q$ is the parameter associated with the line asymmetry. The symmetric Lorentzian profile, i.e. no resonance, is recovered in the limit $|q| \rightarrow \infty$.

The good agreement between the best fit curve and the experimental data out of the DAC (see figures \ref{figura_fano} a,e for MoS$_2$ and WS$_2$ respectively) allows us to adopt a fitting procedure at higher pressures based on the combination of three functions: a polynomial background, a Fano-profile for the E$_{1u}$ phonon, and a sinusoidal curve for the interference fringes (some examples of the fitting procedure are provided in the supporting information). Figures \ref{figura_fano} b,f show the spectra collected within the DAC at selected pressures after the background and fringes subtraction for MoS$_2$ and WS$_2$ respectively.
\\
The parameters obtained from the data fit for the Fano profiles are reported in figure \ref{figura_fano} c,d and in figure \ref{figura_fano} g,h for MoS$_2$ and WS$_2$ respectively as a function of pressure. In both samples, the peak central frequency, $\tilde{\nu}_0$, progressively moves toward higher wavenumbers on increasing pressure. This behaviour is coherent with what has been observed for the Raman-active counterpart of the E$_{1u}$ phonon, that is the E$^1_{2g}$ in-plane mode \cite{Nayak2014, Nayak2015}. In MoS$_2$, the E$_{1u}$ intensity smoothly reduces in the low-pressure range and then rapidly goes to zero in the metallic regime. The screening of the phonon intensity could be regarded as a consequence of the optical response of free electrons, which progressively shield the phonon contribution in the FIR absorption \cite{Stellino2020}. As for the peak profile, figure \ref{figura_fano} d clearly shows that the $q$ parameter keeps almost constant ($q\sim -2.5$) for low-pressure values; then, above P$\sim$18 GPa, just before the spectral weight growth observed before, the modulus of $q$ rapidly increases and the peak symmetrizes. Similarly, in WS$_2$, we observe a nearly constant trend for the peak intensity up to $\sim$15 GPa, followed by an intensity reduction for higher pressure values. Regarding the peak profile, the $q$ parameter maintains a nearly constant value ($q \sim -3$) in the semiconducting regime. Then, just before the onset of the metallic behaviour, the modulus of $q$ abruptly increases leading to the peak symmetrization, in analogy with the MoS$_2$ sample.
\\
Based on the previous discussion, the evolution of the $q$ parameter in the peak profile can be considered a benchmark to evaluate the response of the doping levels to the applied pressure. In particular, the peak symmetrization implies the occurrence of an electronic configuration in which transitions from intra-gap states to conduction band cannot take place. In this perspective, our hypothesis is that the pressure evolution of the sample band structure leads to a progressive reduction of the energy separation between conduction and doping states until the latter ones overlap in correspondence with the metallization pressure $P^{FIR}_{M}$. Once doping levels are hybridized with the conduction band, no more direct electronic transitions from intra-gap states are available for the phonon to couple with, preventing the observation of the Fano resonance in the E$_{1u}$ profiles.

\section{Discussion and conclusion}
\balance
The present work was devoted to the spectroscopic study of the metallic transition in MoS$_2$ and WS$_2$ under pressure in the far infrared range. 
\\
The analysis of the spectral weight under pressure proved itself as a robust, reliable indicator of the onset of the metallic behavior in both compounds, providing us with an estimate of the metallization pressure perfectly compatible with that obtained by transport measurements. The SW trends also gave us a semi-quantitative information on the different rate of increase of the DC conductivity in the two crystals, suggesting a slower transition in WS$_2$ compared with the MoS$_2$ case, in agreement with the results from resistivity measurements. It is worth noticing that a similar analysis conducted on Mo$_{0.5}$W$_{0.5}$S$_2$ \cite{Stellino2021} gave a metallization threshold very close to the average $P^{FIR}_{M}$ of MoS$_2$ and WS$_2$, which is exactly what would be expected for an alloy compound. Moreover, a good correspondence between the metallization onset obtained from pressure-dependent FIR measurements and pressure-dependent transport measurements was found for MoTe$_2$ \cite{Stellino2022}. These results obtained on three different TMD semiconductors and a TMD alloy indicate the SW analysis of the FIR spectrum as a reliable approach for the detection of the metallic behavior in this class of materials. Compared with transport measurements, FIR transmission measurements allow working with constant room temperature and offer the advantage of an easier assembly of the DAC, although they do not enable for a quantitative estimate of physical quantities as the carrier density and mobility.
\\
Regarding the analysis of the phonon modes, the asymmetric profile, shown by the E$_{1u}$ peak at ambient conditions, was modelled in the framework of the Fano theory and interpreted as symptomatic of the coupling between the vibrational mode and a continuum of electronic transitions with comparable energy scales. These transitions cannot originate from a \textit{standard} 

\begin{figure*}[t]
    \centering
    \includegraphics[width=1\textwidth]{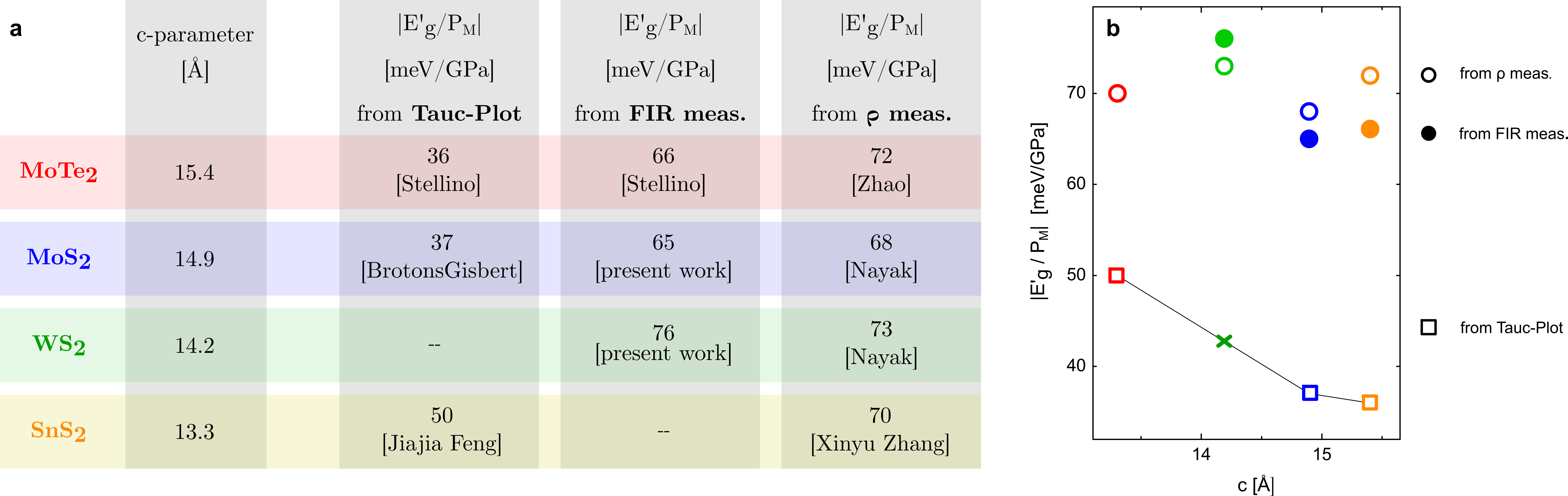}
    \caption{(a) The table reports the estimates of $E'_g / P_{M} $, in which $E'_g $ is the indirect gap energy at ambient pressure and the $P_{M}$ values are the metallization pressures found by Tauc-plot extrapolation in the NIR/visible range \cite{BrotonsGisbert2018, Feng2022, Stellino2022}, by absorption measurements in the FIR \cite{Stellino2022} (and present work), and by transport measurements \cite{Zhao2019, Nayak2014, Nayak2015, Zhang2022}, for different TMD semiconductors. For each sample, we also reported  the c-parameter of the lattice at ambient pressure as listed in \cite{Jain2013}. (b) The figure reports the values of $E'_g / P_{M} $ obtained by Tauc-plot extrapolation in the NIR/visible range (empty squares), by FIR measurements (full circles) and by transport measurements (empty circles) as a function of the c-parameter  for MoTe$_2$ (red), MoS$_2$ (blue), WS$_2$ (green), SnS$_2$ (yellow). The green cross refers to the rough estimate of $E'_g / P_{M} $ for WS$_2$ that we obtained by high-pressure photoluminescence measurements in the range 0-4 GPa (see supporting information).}
    \label{tabella}
\end{figure*}

 excitation from valence to conduction band, which is far too energetic for the phonon to couple with, but they reasonably occur from doping levels located in the proximity of the conduction band minimum. Based on this consideration, the pressure evolution of the Fano resonance can provide us with a unique benchmark to evaluate the response of these doping levels to the applied pressure. In particular, our measurements on MoS$_2$ and WS$_2$ show that the asymmetry parameter extracted from the Fano fit of the phonon line remains finite and nearly constant in the semiconducting regime; then, just before the onset of the metallic behaviour, the peak abruptly symmetrizes suggesting the energy separation between the doping levels and the conduction band minimum goes to zero.
\\
Following the results obtained from the analysis of the Fano resonance in the phonon peaks, we propose an alternative scenario for the metallization process in MoS$_2$ and WS$_2$, in which the metallic behavior at $P^{FIR}_{M}$ is driven by the pressure-induced overlap between doping levels and conduction band minimum, while the electronic band gap is still open. In order to observe the effective closure of the band gap, we would need to further increase the applied pressure, until valence and conduction band cross the Fermi level. This interpretation model, analogous to that proposed for the metallization process in MoTe$_2$ \cite{Stellino2022}, can explain the coincidence between the spectral weight increase and the phonon peak symmetrization under pressure in the measured samples. Moreover, it allows reconciling the apparently controversial results reported in the literature on MoS$_2$, where NIR/visible absorption measurements found a gap closing pressure significantly higher with respect to the metallization pressure obtained by transport measurements and by our FIR measurements. Indeed, absorption spectroscopy in the NIR/visible gives us information about the evolution of the main electronic transition without being affected by the presence of low-energy excitation from extra doping levels. Conversely, transport and FIR measurements are sensible to the increase in the sample conductivity, which might arise independently on the evolution of the valence and conduction band extremes.
\\\\
To further corroborate our hypothesis, we report in figure \ref{tabella}(a) different estimates of $E'_g / P_{M} $, in which $E'_g $ is the indirect gap energy at ambient pressure and the $P_{M}$ values are the metallization pressures found by Tauc-plot extrapolation in the NIR/visible range, by absorption measurements in the FIR, and by transport measurements, for MoTe$_2$, MoS$_2$, WS$_2$ and SnS$_2$.
\\
As shown in figure \ref{tabella}(b), the value of $E'_g / P_{M} $ obtained by Tauc-plot extrapolation steadily increases as the c-parameter of the considered material decreases. The mutual dependence between  $E'_g / P_{M} $ and c supports the idea that the metallization threshold $P_{M}$ found in this case corresponds to the actual gap-closing pressure. Indeed, in TMD semiconductors, the band extremes involved in the indirect gap are dominated by out-of-plane orbitals (like the $p_z$ from the chalcogen atoms) \cite{Cappelluti2013, Kadantsev2012}, which are particularly sensitive to variations in the inter-layer distance. Therefore, the trend of the indirect band gap under pressure is expected to show some kind of correlation with the c-parameter, which defines inter-layer distance at ambient conditions.
\\
In the same perspective, the absence of a clear dependence in the trends of  $E'_g / P_{M} $  \textit{versus} the c-parameter obtained by FIR and transport measurements, very close to each other,  corroborates the hypothesis that the metallic behaviour observed in these experiments is driven by doping effects which are not strictly related to the evolution of the indirect band gap. 
\\\\
To sum up, high pressure FIR measurements on MoS$_2$ and WS$_2$ enabled us to gain a deeper insight into the evolution of the electronic properties of semiconducting TMDs. Based on the obtained results, our idea is that the pressure-induced metallization process in these compounds comprises two distinct charge delocalization processes: the first one, here observed, driven by the reduction in the energy separation between doping levels and conduction band; the second one, taking place at higher pressures, driven by the reduction in the energy separation between valence band maximum and conduction band minimum. The observation of a nearly identical phenomenology in both MoS$_2$ and WS$_2$, not to mention the similar results obtained on MoTe$_2$ in the literature, might indicate that the presence of unintentional doping levels plays a key role in the pressure evolution of the electronic properties of most of semiconducting TMDs and thus must be carefully considered in all experimental works aiming at studying metallization processes in this class of materials.

%\bibliographystyle{unsrt}
%\bibliography{mybib}

\end{document}